\documentstyle[preprint,aps]{revtex}

\begin{document}
\title{The RR interval spectrum, the ECG signal and aliasing}
\author{A. Gersten$^{(1),(4)}$, O. Gersten$^{(3)}$, A. Ronen$^{(2)}$and Y. Cassuto$%
^{(2),(4)}$}
\address{$^{(1)}$Dept. of Physics, $^{(2)}$Dept. of Life Sciences,\\
$^{(3)}$Dept. of Mathematics and Computer Sciences, \\
$^{(4)}$Unit of Biomedical Engineering,\ \\
Ben-GurionUniversity of the Negev}
\date{August 25, 1999}
\maketitle
\pacs{87.19.Hh, 87.80.Tq}

\begin{abstract}
We discuss the relationship between the RR interval spectral analysis and
the spectral analysis of the corresponding ECG signal from which the RR
intervals were evaluated. The ECG signal spectrum is bounded below the
frequency f$_{B}$ by using an electronic filter and sampled at rate larger
than 2f$_{B}$, thus excluding aliasing from spectral analysis. A similar
procedure cannot be applied to the RR interval spectral analysis, and in
this case aliasing is possible. One of our main effort in this paper is
devoted to the problem of how to detect aliasing in the heart rate spectral
analysis. In order to get an insight we performed an experiment with an
adult man, in which the ECG signal was detected in a case where the
breathing rate was larger than half the heart rate. A constant breathing
rate for time intervals exceeding 5 minutes was monitored with good \
accuracy using a special breathing procedure. The results show distinctively
a very sharp peak in the spectral analysis of the ECG signal and
corresponding (diffused) aliasing peaks in the RR interval spectral analysis.

New method of dealing with unevenly sampled data was developed which has
interesting anti-aliasing properties. There are indications that the VLF
peaks of the RR spectrum are originated by aliasing. Some of the LF peaks
may have the same property.

\medskip 

{\bf Keywords:} Hart rate, ECG signal, Spectral analysis, Aliasing
\end{abstract}

\section{Introduction}

The R-R interval spectral analysis is usually based on heart rate data
collected in two ways. In one method the data are collected by analog to
digital conversion of the ECG signal and computer evaluation of the R-R
intervals from the ECG signal. In the second method, devices are used whose
output is the R-R interval alone. The advantage of the first method is the
control of accuracy and flexibility of the evaluations. The second method
has the advantage of storing smaller amount of data, and it can be easily
used on-line.

\ \ \ In the first method, usually the number of collected data (sampled ECG
signal) is of two to three orders of magnitude larger than the R-R interval
data. Thus if only R-R interval is analyzed a large amount of data is
unused. In this paper we are trying to take advantage of the ECG sampled
signal and to derive new information in addition to the conventional R-R
interval analysis \cite{sayers},,\cite{kamath},\cite{malik}, \cite{malik2}.

\ \ \ The ECG signal spectrum is bounded below the frequency f$_{B}$ by
using an electronic filter and sampled at rate larger than 2f$_{B}$, thus
excluding aliasing from spectral analysis.\cite{oppen} A similar procedure
cannot be applied to the R-R interval spectral analysis, and in this case an
aliasing is possible. One of our main efforts in this paper is devoted to
the problem of how to detect aliasing in the R-R interval spectral analysis.

\ \ \ \ \ In order to get an insight, we performed an experiment, in which
the ECG signal of one of the authors (A.G) was detected while the breathing
rate was larger than half the heart rate. A constant breathing rate for a
time exceeding 5 minutes was monitored with good \ accuracy using a special
breathing procedure with a metronome. The results show distinctively a very
sharp peak in the spectral analysis of the ECG signal and corresponding
(diffused) aliasing peaks in the R-R interval spectral analysis.

The spectral analysis of the ECG signal was performed with the standard FFT
procedures. The spectral analysis of the R-R intervals was performed with
several techniques in order to take into consideration that the data were
unevenly sampled. This is presented in section 2. In section 3 we discuss
the possibility of aliasing in the spectral analysis of the R-R intervals.
In section 4 we compare power estimations of ECG's and R-R intervals of 3
experiments. In section 5 we analyze the results. In section 6 summary and
conclusions are presented.

\section{Spectral Analysis of Unevenly Sampled Data}

The methods of spectral analysis are well developed for evenly sampled data 
\cite{oppen},\cite{cohen}. The R-R interval data are unevenly sampled in
time. In most cases an analysis is performed with respect to beat numbers
which are evenly spaced. We will below justify this method using least
squares principles. But as was recently indicated by Laguna et Al. \cite
{laguna}, the resampling of data is causing the appearance of additional
harmonics. They recommend to use a method developed by N. R. Lomb \cite{lomb}%
. The errors of resampling the beats can, to large extent, be overcome by
using a cubic spline interpolation. In this work we are suggesting a new
method of treating unevenly sampled data, which, unexpectedly, gave good
results beyond the Nyquist frequency.

\subsection{Analysis according to beat numbers}

Let us assume that the RR intervals are given at unevenly sampled times $%
t_{n}$, with the values $s\left( t_{n}\right) ,$ where $n$ is the beat
number, $n=1\cdots N$. Let us divide the interval $\left[ t_{1},t_{N}\right] 
$ into equal subintervals

\begin{equation}
\Delta \tau =\frac{t_{N}-t_{1}}{N-1},  \label{1}
\end{equation}
and let us generate in the interval $\left[ t_{1},t_{N}\right] $ evenly
sampled times:

\begin{equation}
\tau _{n}=(n-1)\Delta \tau +t_{1}.  \label{2}
\end{equation}

\bigskip We will use the discrete time Fourier transform (DFT) for a basis
formed from the evenly sampled times $\tau _{n}$. We will assume that

\begin{equation}
s(t_{n})=\frac{1}{N}\sum_{k=1}^{N}S_{k}\exp \left( i\omega _{k}\tau
_{n}\right) ,\qquad \omega _{k}=2\pi \left( k-1\right) /\left( N\Delta \tau
\right) .  \label{3}
\end{equation}

The coefficients $S_{k}$ will be determined by minimizing the expression

\begin{equation}
\sigma =\sum_{n=1}^{N}\left\{ \left[ s(t_{n})-\frac{1}{N}\sum_{k=1}^{N}S_{k}%
\exp \left( i\omega _{k}\tau _{n}\right) \right] \left[ s(t_{n})-\frac{1}{N}%
\sum_{k=1}^{N}S_{k}^{\ast }\exp \left( -i\omega _{k}\tau _{n}\right) \right]
\right\}  \label{4}
\end{equation}

with the result

\begin{equation}
S_{k}=\sum_{n=1}^{N}s\left( t_{n}\right) \exp \left( -i\omega _{k}\tau
_{n}\right) .  \label{5}
\end{equation}

Eqns. \ref{5} and \ref{3} can be handled easily with standard FFT programs.
This is the usual procedure which is adopted in most of the papers dealing
with R-R interval analysis.\cite{malik}\cite{malik2}

\subsection{Other Methods}

FFT can be applied more efficiently if the unevenly sampled data are
interpolated at evenly spaced intervals of Eq. \ref{2}. The cubic spline
interpolation is one of the good ways to do it.

The Lomb method \cite{lomb} was extensively analyzed in ref. \cite{laguna}.
We give here only the formulae in the form of the Lomb normalized periodogram

\begin{equation}
P_{X}\left( \omega _{k}\right) =\frac{1}{2\sigma ^{2}}\left\{ \frac{\left[
\sum_{n=1}^{N}\left[ s\left( t_{n}\right) -\bar{s}\right] \cos \left( \omega
_{k}\left( t_{n}-\tau \right) \right) \right] ^{2}}{\sum_{n=1}^{N}\cos
^{2}\left( \omega _{k}\left( t_{n}-\tau \right) \right) }+\frac{\left[
\sum_{n=1}^{N}\left[ s\left( t_{n}\right) -\bar{s}\right] \sin \left( \omega
_{k}\left( t_{n}-\tau \right) \right) \right] ^{2}}{\sum_{n=1}^{N}\sin
^{2}\left( \omega _{k}\left( t_{n}-\tau \right) \right) }\right\}  \label{6}
\end{equation}

where $\bar{s}$ and $\sigma ^{2\text{ \ \ }}$are the mean and variance of
the data and the value of $\tau $ is defined as

\begin{equation}
\tan \left( 2\omega _{k}\tau \right) =\frac{\sum_{n=1}^{N}\sin \left(
2\omega _{k}t_{n}\right) }{\sum_{n=1}^{N}\cos \left( 2\omega
_{k}t_{n}\right) }  \label{7}
\end{equation}

\subsection{Non-Uniform Discrete Fourier Transform (NUDFT)}

We present here a new method of treating unevenly spaced events which we
call the ''non-uniform discrete Fourier transform'' (NUDFT).

Let us assume that $s\left( \tau _{n}\right) $ are the exact values of the
signal at the points given by Eq. \ref{2}. The corresponding DFT is

\begin{equation}
S_{k}=\sum_{n=1}^{N}s\left( \tau _{n}\right) \exp \left( -i\omega _{k}\tau
_{n}\right) .  \label{8}
\end{equation}

Our aim is to find a good approximation to this expression in terms of the
unevenly sampled signal $s(t_{n})$.

We start with the Euler summation formula

\begin{equation}
\sum_{n=1}^{N}f\left( \tau _{n}\right) =\frac{1}{\Delta \tau }\int_{\tau
_{1}}^{\tau _{N}}f\left( \tau \right) d\tau +\frac{1}{2}\left[ f\left( \tau
_{1}\right) +f\left( \tau _{N}\right) \right] +\frac{\Delta \tau }{12}\left[
f^{\prime }\left( \tau _{N}\right) -f^{\prime }\left( \tau _{1}\right) %
\right] +O(\Delta \tau ^{2})  \label{9}
\end{equation}
and make the following decomposition of the integral on the right hand side
of Eq.\ref{9}:

\begin{equation}
\int_{\tau _{1}}^{\tau _{N}}f\left( \tau \right) d\tau
=\int_{t_{1}}^{t_{2}}f\left( \tau \right) d\tau +\int_{t_{2}}^{t_{3}}f\left(
\tau \right) d\tau +\cdots +\int_{t_{N-1}}^{t_{N}}f\left( \tau \right) d\tau
\label{10}
\end{equation}

and approximate each of the integrals on the right hand side with the
trapezoidal rule:

\begin{equation}
\int_{\tau _{1}}^{\tau _{N}}f\left( \tau \right) d\tau =\frac{1}{2}\left[
f\left( t_{1}\right) +f\left( t_{2}\right) \right] \left( t_{2}-t_{1}\right)
+\cdots +\frac{1}{2}\left[ f\left( t_{N-1}\right) +f\left( t_{N}\right) %
\right] \left( t_{N}-t_{N-1}\right) +O\left( \Delta \tau \right)  \label{11}
\end{equation}

From Eqs. \ref{9} and \ref{11} we obtain:

\[
\sum_{n=1}^{N}f\left( \tau _{n}\right) =\frac{1}{2\Delta \tau }\left\{ \left[
f\left( t_{1}\right) +f\left( t_{2}\right) \right] \left( t_{2}-t_{1}\right)
+\cdots +\left[ f\left( t_{N-1}\right) +f\left( t_{N}\right) \right] \left(
t_{N}-t_{N-1}\right) \right\} 
\]

\begin{equation}
+\frac{1}{2}\left[ f\left( t_{1}\right) +f\left( t_{N}\right) \right]
+O\left( \Delta \tau \right) .  \label{12}
\end{equation}

When the $t_{n}$ are equally spaced Eq. \ref{12} becomes an identity with
the $O\left( \Delta \tau \right) =0$, therefore it seems to us that Eq. \ref
{12} is satisfied with an higher accuracy than just $O\left( \Delta \tau
\right) .$ Eq. \ref{12} can be applied to approximate Eq. \ref{8} with the
substitution

\begin{equation}
f\left( t_{n}\right) =s\left( t_{n}\right) \exp \left( -i\omega
_{k}t_{n}\right) ,  \label{13}
\end{equation}
and the final result, the approximation to Eq. \ref{8}, after rearranging
the terms, becomes:

\begin{equation}
S_{k}=\sum_{n=1}^{N}c_{n}s\left( t_{n}\right) \exp \left( -i\omega
_{k}t_{n}\right) +O\left( \Delta \tau \right) ,  \label{14}
\end{equation}

where

\begin{equation}
\begin{array}{c}
c_{1}=\frac{\Delta \tau +t_{2}-t_{1}}{2\Delta \tau }, \\ 
c_{2}=\frac{t_{3}-t_{1}}{2\Delta \tau }, \\ 
\vdots \\ 
c_{N-1}=\frac{t_{N}-t_{N-2}}{2\Delta \tau }, \\ 
c_{N}=\frac{\Delta \tau +t_{N}-t_{N-1}}{2\Delta \tau },
\end{array}
\label{15}
\end{equation}

with the inverse formula 
\begin{equation}
s(\tau _{n})=\frac{1}{N}\sum_{k=1}^{N}S_{k}\exp \left( i\omega _{k}\tau
_{n}\right) +O\left( \Delta \tau \right) ,\qquad \omega _{k}=2\pi \left(
k-1\right) /\left( N\Delta \tau \right) ,  \label{16}
\end{equation}

which is an interpolation formula for $s\left( t_{n}\right) $ at the evenly
spaced points $\tau _{1}\cdots \tau _{N}$.

\section{\protect\bigskip Aliasing}

Aliasing is a result of undersampling and is a well known phenomenon. In
ref. \cite{gersten} aliasing was looked upon from the point of view of
symmetry. It is an example of wrong symmetry, and as such should be given
more attention. It is the outcome of an incomplete basis. It was found in
ref.\cite{gersten} , that for evenly sampled data with a sampling rate $%
f_{S} $, the spectral amplitude $S(f)$ evaluated with FFT, has the following
symmetry properties

\begin{equation}
\left| S\left( f\right) \right| =\left| S\left( f\pm f_{S}\right) \right|
=\left| S\left( -f\pm f_{S}\right) \right| =\left| S\left( \pm f\pm
nf_{S}\right) \right| ,  \label{17}
\end{equation}
where $f$ is the frequency and n is an arbitrary integer.

In order to avoid the aliasing symmetry of Eq. \ref{17}, the frequencies
should be bounded by the Nyquist frequency (denoted here by $f_{B}$)
according to 
\begin{equation}
f_{B}=\frac{f_{S}}{2}.  \label{19}
\end{equation}

The ECG signal was sampled with sampling rate 250 Hz, and an electronic
filter was applied, which have eliminated practically all frequencies above
32 Hz, thus aliasing can not occur at frequencies below 125 Hz or even below
32 Hz. The R-R intervals were calculated directly from the ECG signal. The
sampling rate for R-R intervals can be defined only for evenly sampled data,
for the methods which interpolates the unevenly sampled data, or one can
consider the average sampling rate from Eq. \ref{1}, in both cases

\begin{equation}
\bar{f}_{S}=1/\Delta \tau =2f_{N},  \label{18}
\end{equation}
where $f_{N}$ is the Nyquist frequency for the R-R intervals. As the ECG
signal contains frequencies much grater than $f_{N}$, and the R-R intervals
are derived from the ECG signal, one can not be sure that the spectral
analysis of the R-R intervals is free from aliasing. As a matter of fact
there are indications of aliasing in some rare cases.\cite{witte}\cite
{rother}\cite{nilsson}\cite{zwiener}\cite{zwiener1}\cite{zwiener2} One way
to identify aliasing is to change the sampling rate and follow the changes
in the spectrum. Unfortunately, for the R-R intervals, one can not speak
about a definite sampling rate, but rather can consider a distribution of
sampling rates. The changes in sampling rate required to observe aliasing
are of the same order as the fluctuations in the sampling rate. Therefore in
practice it is almost impossible to observe consistent changes in the
spectrum slightly changing the heart rate.

Other possibility of detecting aliasing is by comparing the heart \ rate
spectrum with the ECG signal spectrum. Marked differences below the Nyquist
frequency for the power distribution of the RR intervals compared to the ECG
signal power distribution in the same range may indicate aliasing. But we do
not have yet a sound basis to treat this problem.

We have devised an experiment which definitely demonstrates the aliasing in
the R-R intervals spectrum. To the best of our knowledge this is the first
experiment in which one can exactly know the correct frequency above the
Nyquist frequency and can follow the development of the aliasing, which
appears to be diffused to great extent because the symmetry of Eq. \ref{17}
is represented not by one sampling rate but by a distribution of sampling
rates, as the R-R interval is unevenly sampled.

Below we describe 3 experiments. One of them was devised to demonstrate
aliasing and the other two for learning about the relations between the R-R
interval spectrum and the spectrum of the ECG signal.

\section{Three Experiments}

We present below results of three experiments. In the first experiment the
ECG signal was collected in a normal resting state. The aim of this
experiment was to compare the ECG spectrum with the R-R intervals spectrum.
In the second experiment very slow breathing was monitored at a rate of 0.04
Hz. Again the ECG and R-R interval spectra were compared. In the third
experiment very fast breathing was accurately monitored at the rate of
74/min and 84/min. These respiratory rates were above half of the heart
rates thus allowing to observe in detail the development of aliasing.

\subsection{The First Experiment}

In this experiment (linked with the names of Zahi and Ori, where the second
is one of the authors: O.G) which was done in normal, resting conditions, we
compare the power estimation of the R-R interval and the ECG signal, from
which the R-R interval was obtained. The ECG signal was sampled at a rate of
250 Hz. Stable intervals of 7 minutes duration were chosen for analysis.

In Fig. 1a the power distribution of the ECG signal of Zahi is depicted. The
attenuation of the power with increasing frequency above 12 Hz is due to the
action of an electronic filter. Above 32 Hz the contribution is practically
zero. The average heart rate was 0.97 Hz. The above results were zoomed to
the interval [0-12] Hz in Fig. 1b. One can see distinctively the peak around
the average heart rate and the higher harmonics of this peak. The second
harmonic is missing, but the third, fourth, fifth and sixth are
distinctively visible, higher harmonics became more and more smeared and
indistinguishable above the sixth harmonic. One should also note the large
difference in power in the heart rate range, below the Nyquist frequency of
0.49Hz, which is much smaller compared to the peak around the average heart
rate (0.97 Hz).

The power distribution of the RR intervals in the range [0-0.5] Hz was
computed according to the methods discussed in section 2 and are presented
in Figs. 2a (DFT, beat number analysis), 2b (Spline interpolation), 2c
(NUDFT). For comparison also the power distribution of the ECG signal in the
above range is presented in Fig. 2d.

The results of Figs. 2a, 2b, and 2c are quite similar, but the spline
interpolation (Fig. 2b) and the NUDFT (Fig. 2c) are practically identical.
The three graphs show the structure commonly found in the power estimation
analysis of RR intervals, namely the existence of the ''high frequency''
(HF), ''low frequency'' (LF) and \ the ''very low frequency'' (VLF) peaks.
The ECG spectrum shows qualitatively the same structure\ (but not a
quantitative agreement), except that the ECG spectrum is highly suppressed
below 0.04 Hz, in the VLF region, indicating a possibility of aliasing in
this region in the RR analysis.

In Figs. 3 and 4a-4d the results of Ori are presented. The conclusions are
similar to those of Zahi, except that in the ECG spectrum both VLF and LF
peaks are missing, indicating the possibility of aliasing in these regions
for the RR analysis. Also in the ECG spectrum of Ofek, Fig. 5b the VLF and
LF, present in Fig. 5a, are missing. VLF is missing in J.C.'s ECG spectrum
(see Figs. 6a-6b).

\subsection{The Second Experiment}

In this experiment (linked again with the name Ori) we have checked the ECG
spectrum near the VLF region, as the VLF was absent in the ECG spectrum for
the resting state in the first experiment. The question was whether such a
result persists in all ECG spectra. Therefore we have probed the VLF region
by monitoring very prolonged breathing with a rate of 0.04 Hz. For the
spectrum of RR intervals we found that the DFT, Spline interpolation and
NUDFT gives similar results, and again NUDFT was practically identical to
the spline interpolation. Therefore we present only the results of NUDFT,
which are presented in Fig. 7a. For comparison the spectrum of the ECG
signal is given in Fig. 7b. In Fig. 7a one can see a very clean pattern of a
peak at 0.04 Hz and its higher harmonics. In Fig. 7b one can see a similar
but somewhat diffused pattern. Thus this experiment indicates that similar
respiratory patterns exists in both the RR as well as in the ECG signal.

\subsection{The Third Experiment}

In this experiment (linked to the name Alex, who is one of the authors: A.G)
very fast breathing was accurately monitored at the rate of 74/min and
84/min respectively. These rates were well above half of the average heart
rate thus allowing to observe in detail the development of aliasing. In Fig.
8 the ECG spectrum is dominated by the very high and narrow peak at the
frequency $f_{1}=1.234Hz$, also its higher harmonics can be distinctively
seen. The frequency $f_{1}$ is just the breathing frequency 74/min. In the
same figure one can also see the diffused peaks near the average heart rate
frequency of 1.636 Hz \ and its higher harmonics. One should observe
aliasing at about $1.636-$ $f_{1}=0.402Hz$. Indeed one can see diffused
peaks around that frequency in Fig. 9a, which displays the power estimation
of the RR intervals using the NUDFT (which below the Nyquist rate is similar
to the spline interpolation). The width of this region can be estimated by
noting that the RR intervals have different instantaneous sampling rates
which are equal to the inverse of the RR interval time. In Fig. 10 we have
calculated the distribution of the sampling rates by dividing the frequency
region into 100 beans. We have shifted that distribution by subtracting $%
f_{1}$. As one can see the results are confined approximately to the region
0.32-0.47 Hz. Indeed the aliasing peaks of Fig. 9a appear in this region.
The pictures below the Nyquist frequency are very similar for the DFT,
NUDFT, the spline interpolation and the Lomb method (Fig. 9b) with a similar
aliasing behavior.

In principle the NUDFT and the Lomb methods should not be used above the
Nyquist frequency. Surprisingly enough we have found that both methods have
a sharp peak at $f_{1}$, as can be seen in Figs. 9a and 9b. Both methods do
not have the aliasing symmetry of the DFT as given by Eq. \ref{17},
therefore the results are not symmetric with respect to the Nyquist
frequency (half the sampling rate), as it is satisfied, for example, in the
case of the spline interpolation. We have found an exact result at $f_{1}$
and a diffused aliasing around 0.4 Hz. It is interesting to note that both
methods give almost the same result below and above the Nyquist frequency.
One can interpret the appearance of the sharp peak at $f_{1}$ as a result of
a partial destruction of aliasing symmetry due to uneven samplings.

Similar results for the breathing frequency 84/min are presented in Figs.
11-12.

\section{Summary and Conclusions}

\ \ \ The ECG signal spectrum is bounded below the Nyquist frequency f$_{B}$
by using an electronic filter and sampled at rate larger than 2f$_{B}$, thus
excluding aliasing from spectral analysis. A similar procedure cannot be
applied to the RR interval spectral analysis, and in this case an aliasing
is possible. One of our main efforts in this paper was devoted to the
problem of how to detect aliasing in the R-R interval spectral analysis.

In order to get insight into this problem three experiments have been
analyzed. In the first experiment the ECG signal was collected in a normal
resting state. The aim of this experiment was to compare the ECG spectrum
with the R-R interval spectrum. In the second experiment very slow breathing
was monitored at a rate of 0.04 Hz. Again the ECG and R-R interval spectra
were compared. In the third experiment very fast breathing was accurately
monitored at the rate of 74/min and 84/min respectively. These respiratory
rates were above half of the heart rates thus allowing to observe in detail
the development of aliasing.

The experiments which were described above led us to the following
conclusions:

\begin{enumerate}
\item  The spectral analysis of the ECG signal is more sensitive and
accurate compared to the R-R interval spectral analysis and is free from
aliasing. Still in the present stage it contains too much information to be
of practical use. Efforts should be made to understand what will be the best
way to extract information (not related to the heart condition alone as in
the standard analysis of ECG) about the external influences on the heart
signal.

\item  We have conducted an experiment which gave a clear insight about the
mechanism of aliasing in the R-R interval spectrum. The very sharp peak in
the spectrum of the ECG signal, which came as the result of enforced quick
breathing, reappeared as a diffused signal in the RR spectrum. The extension
of the diffuseness agrees with the extension of the sampling rates of
unevenly sampled data..

\item  The VLF peak observed in the R-R interval spectrum is usually missing
in the ECG spectrum. This lead us to suspect that the VLF observed in the RR
spectrum has its origin in aliasing.

\item  In some cases the LF peak does not show up in the ECG spectrum. This
led us to suspect that part of the LF peak is of aliasing origin.

\item  Unlike in electronic devices, it is very difficult to devise
procedures to detect aliasing in humans. In electronic devices aliasing can
be easily detected by changing the sampling rate. In humans the fluctuations
of the heart rate are of the same order as the required changes in the
sampling rates. It will be an important task to develop a proper procedure
for detecting aliasing in humans.

\item  We have developed a new technique for spectral analysis for unevenly
sampled data called non-uniform discrete Fourier transform (NUDFT). When
employed to the RR data, below the Nyquist frequency, it gave similar
results as those obtained by interpolating the data with a cubic spline.
Above the Nyquist frequency, the correct peak in the spectrum was detected
with great accuracy. A similar result was obtained with the recently
rediscovered Lomb method. We interpret this unexpected result by a partial
destruction of aliasing symmetry in both methods. More efforts should be
made in order to understand the anti-aliasing properties of the above
methods.

\item  We consider aliasing to be a wrong symmetry, resulting from the use
of an incomplete basis, which has intrinsic symmetries inconsistent with the
properties of the signal. Aliasing can be partially removed by reducing the
symmetry of the basis.
\end{enumerate}

\newpage

\ \ \ \ \ \ \ \ \ \ \ \ \ \ \ \ \ \ \ \ \ \ \ \ \ \ \ \ \ \ \ \ \ \ \ \ \ \
\ \ \ \ \ \ {\Large Figure Captions}

\begin{itemize}
\item  Figure 1. The relative power of the ECG signal of Zahi, a) in the
spectral range of 0-36 Hz, b) in the spectral range of 0-12 Hz.

\item  Figure 2. The relative power computed (from the ECG signal of Zahi)
by four different methods, in the spectral range of 0-0.5 Hz, a) by DFT, b)
by spline interpolation of the RR data, by NUDFT, d) from the ECG signal.

\item  Figure 3. The relative power of the ECG signal of Ori.

\item  Figure 4. The relative power computed (from the ECG signal of Ori) by
four different methods, in the spectral range of 0-0.52 Hz. a) by DFT, b) by
spline interpolation of the RR data, c) by NUDFT, d) from the ECG signal.

\item  Figure 5. The relative power computed (from the ECG signal of Ofek)
by two different methods, in the spectral range of 0-0.6 Hz, a) by spline
interpolation of the RR data, b) from the ECG signal.

\item  Figure 6. The relative power computed (from the ECG signal of J.C.)
by two different methods, in the spectral range of 0-0.46 Hz, a) by spline
interpolation of the RR data, b) from the ECG signal.

\item  Figure 7. The relative power computed (from the ECG signal of Ori
with breathing rate of 0.04 Hz) by two different methods, in the spectral
range of 0-0.62 Hz, a) by NUDFT, b) from the ECG signal.

\item  Figure 8. The relative power of the ECG signal of Alex with a
breathing rate of 1.234 Hz.

\item  Figure 9. The relative power computed (from the ECG signal of Alex
with a breathing rate of 1.234 Hz) by two different methods, in the spectral
range of 0-1.5 Hz, a) by NUDFT, b) from the ECG signal.

\item  Figure 10. A 100 bin histogram of the heart rates of Alex which are
subtracted by the breathing rate of 1.234 Hz.

\item  Figure 11. The relative power of the ECG signal of Alex with a
breathing rate of 1.404 Hz.

\item  Figure 12. The relative power computed (from the ECG signal of Alex
with a breathing rate of 1.404 Hz) by two different methods, in the spectral
range of 0-1.6 Hz, a) by NUDFT, b) from the ECG signal.
\end{itemize}

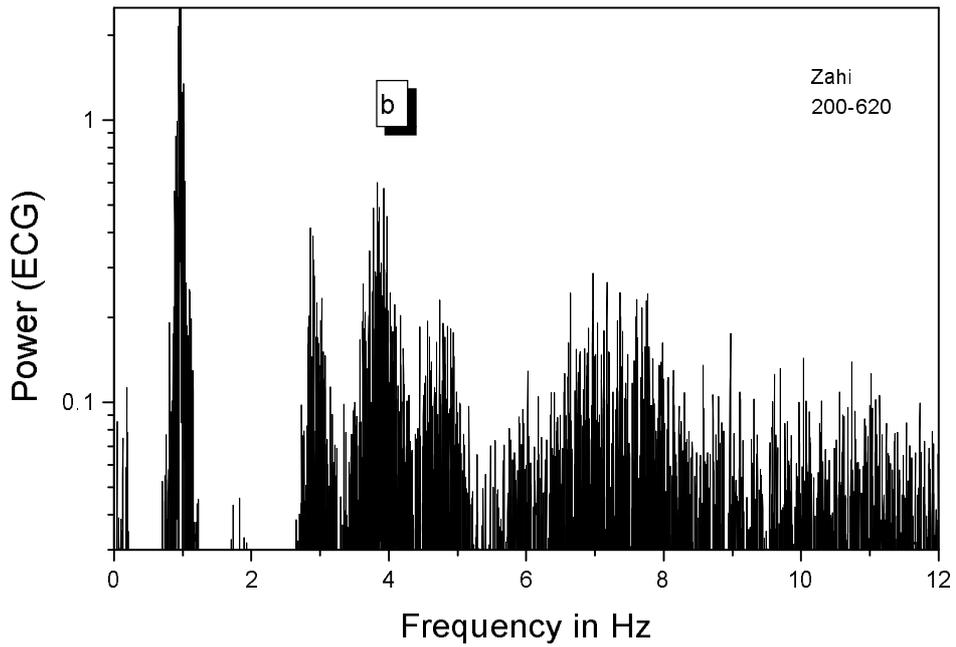
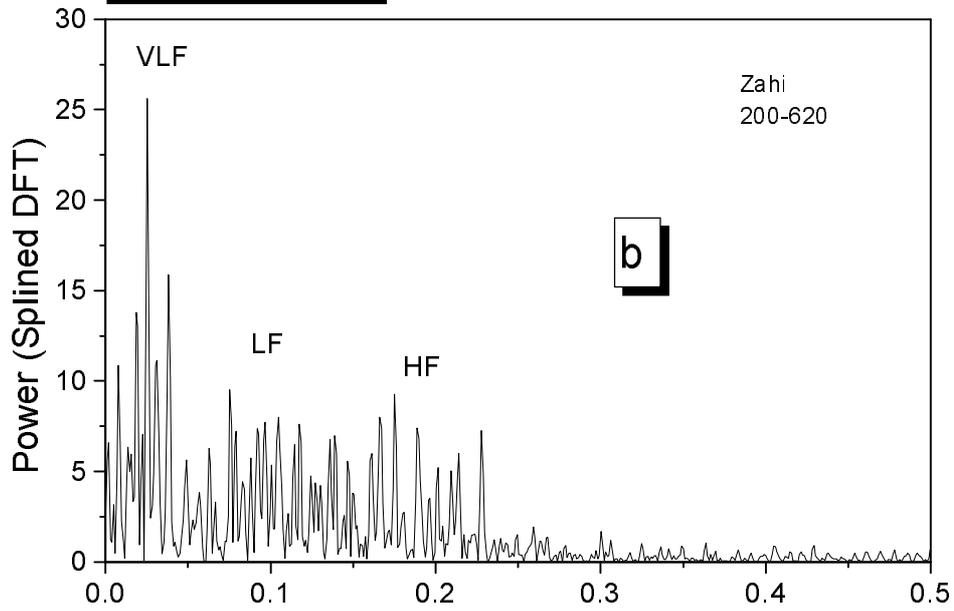
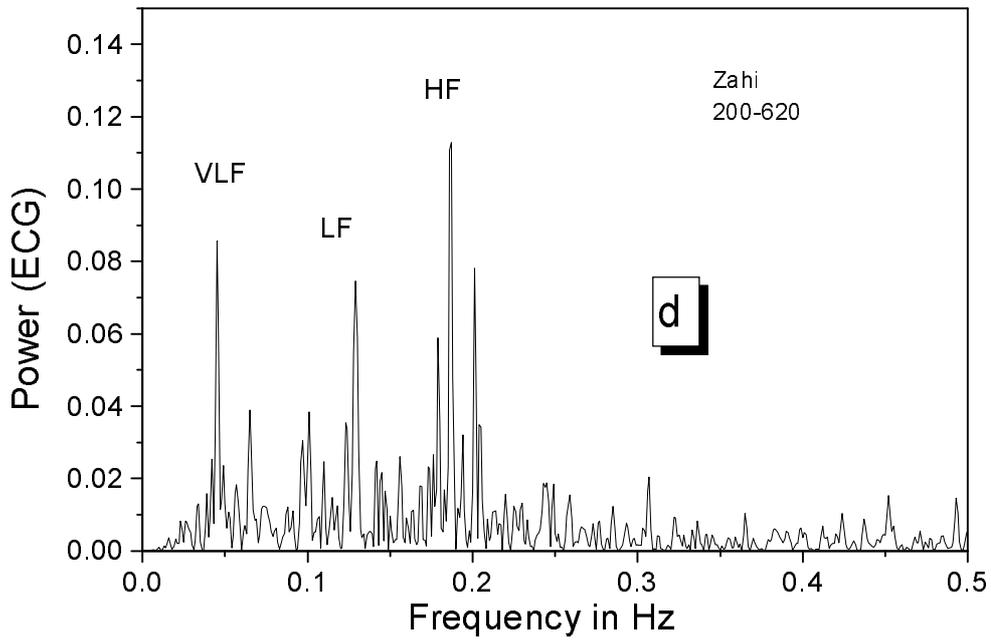
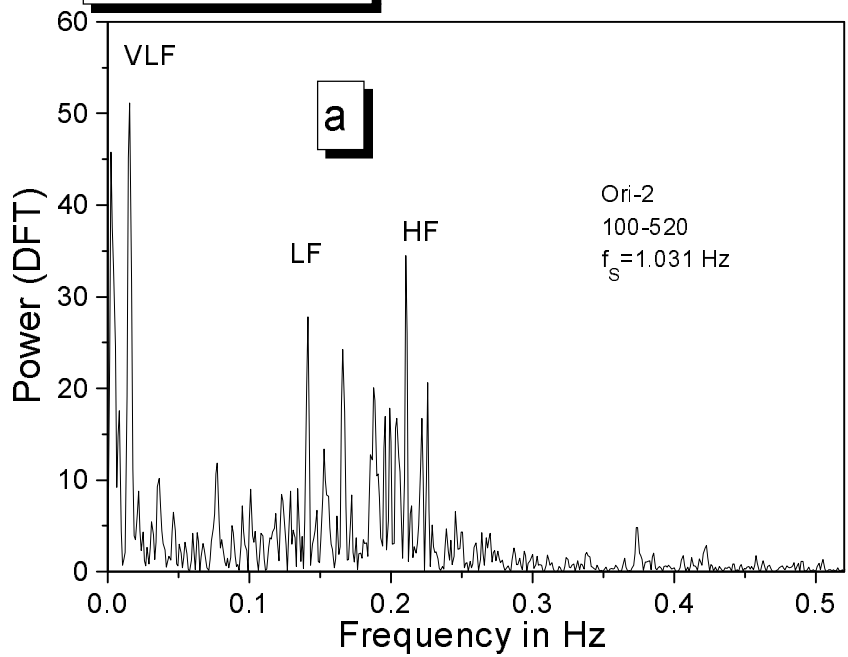
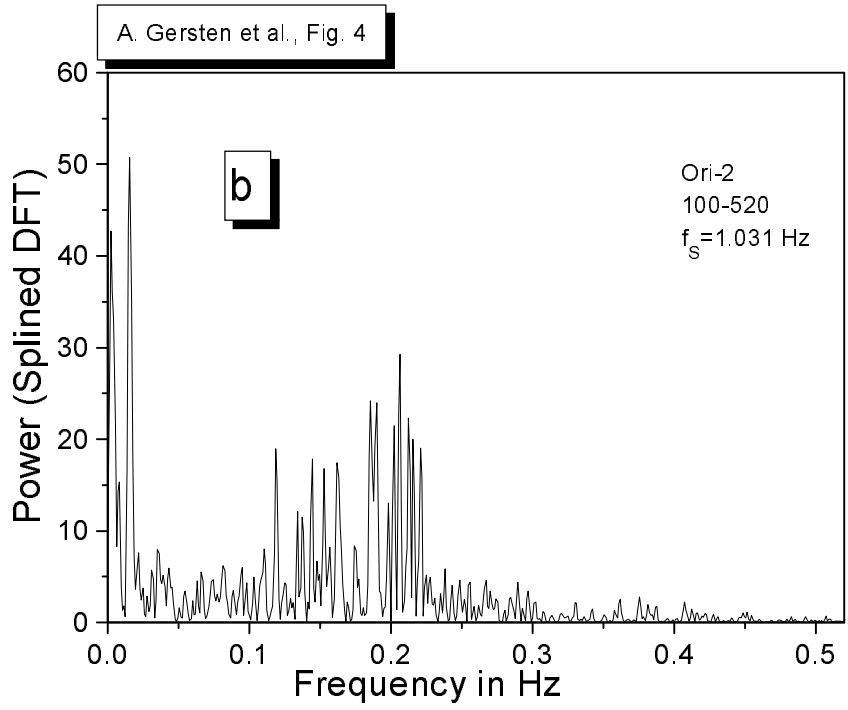
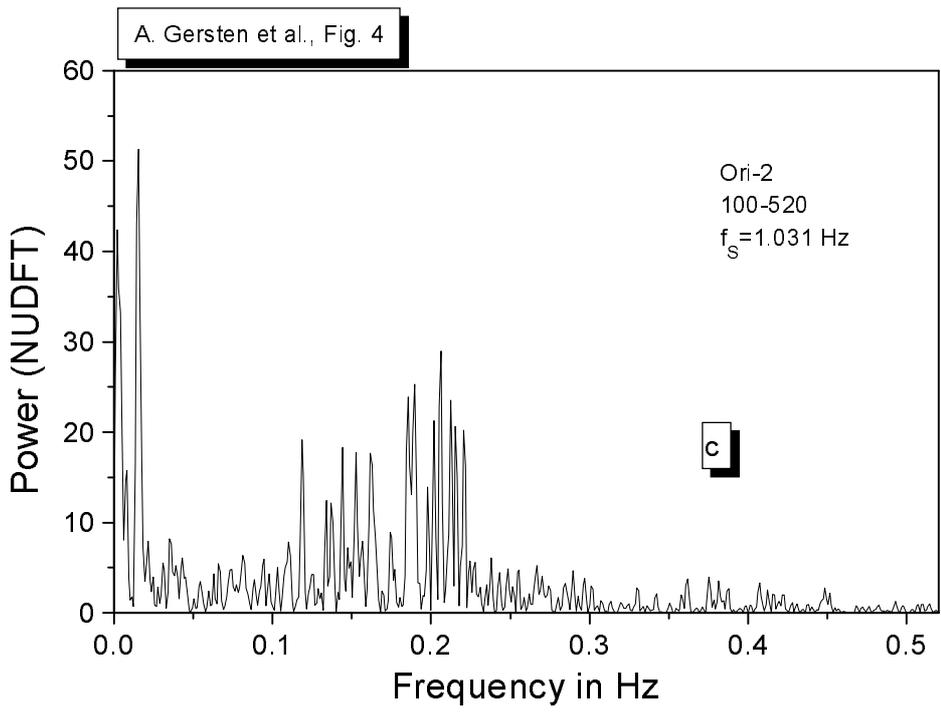
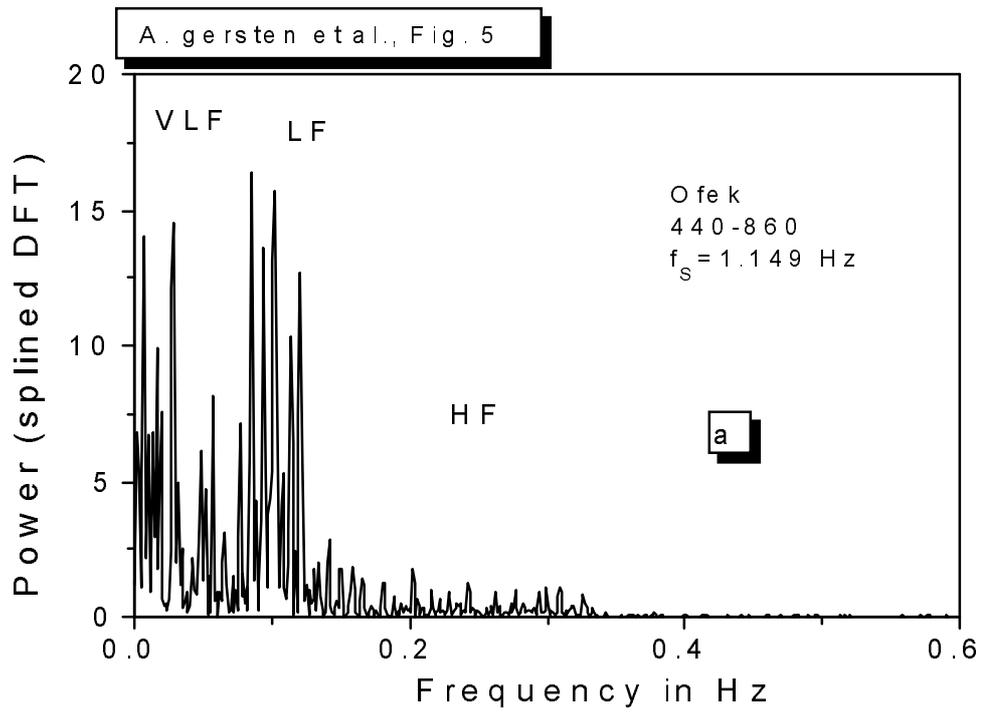
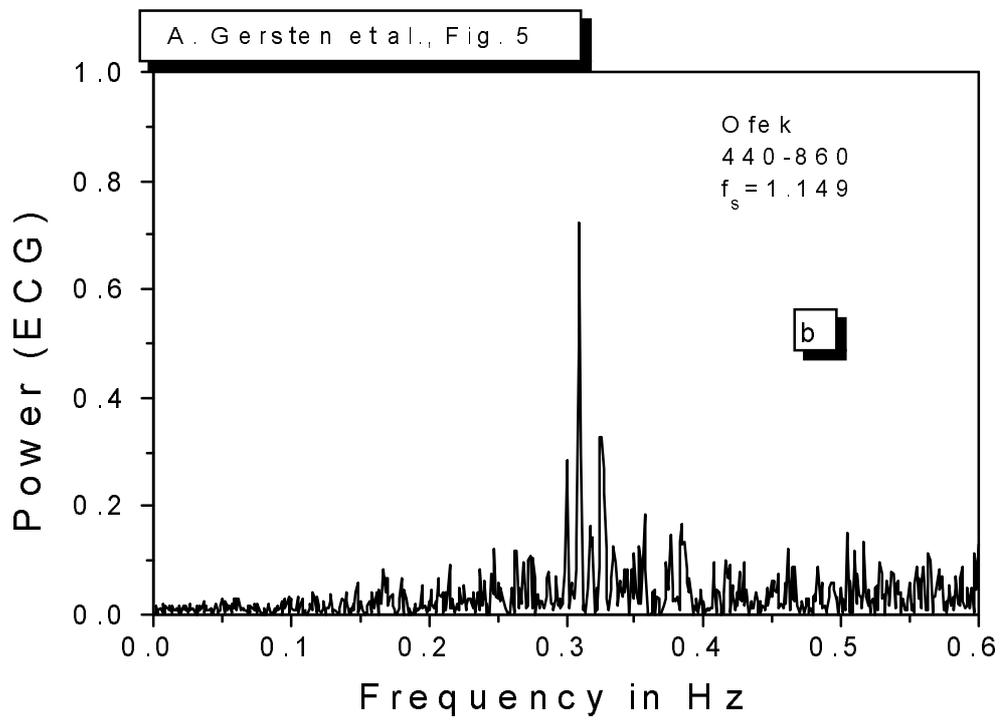
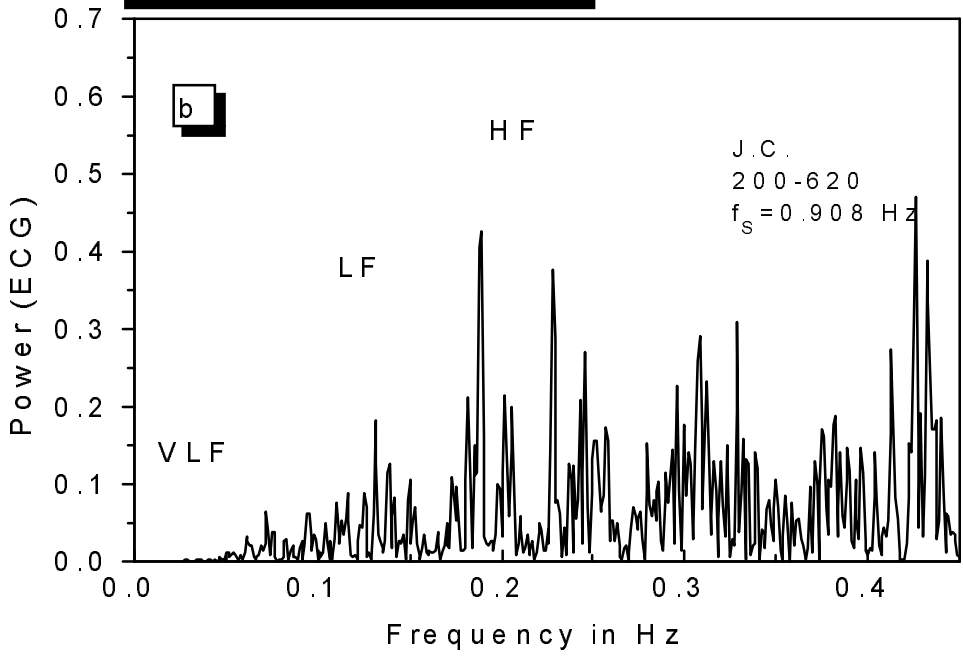
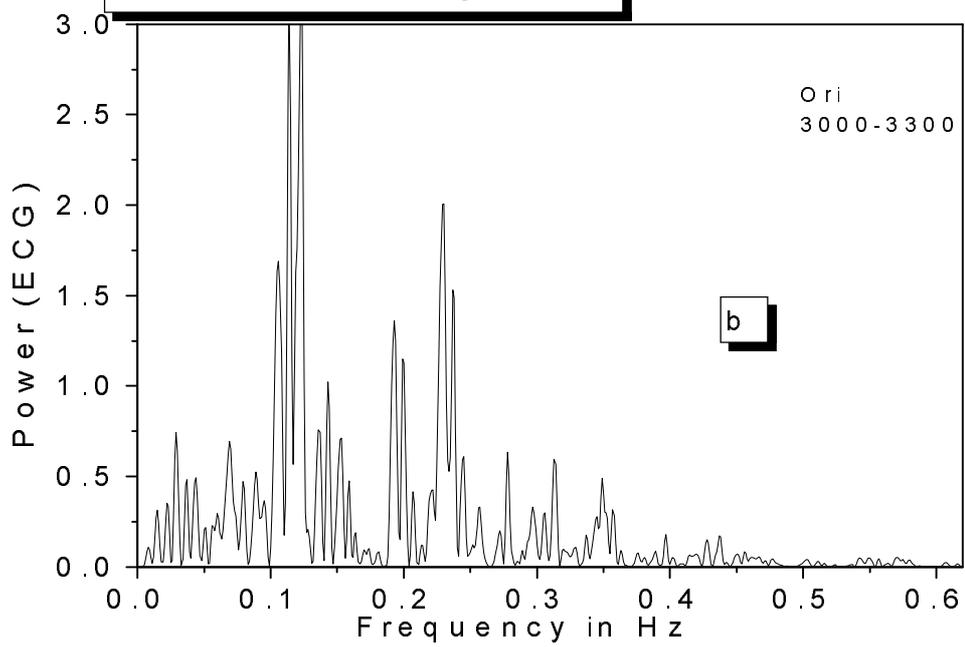
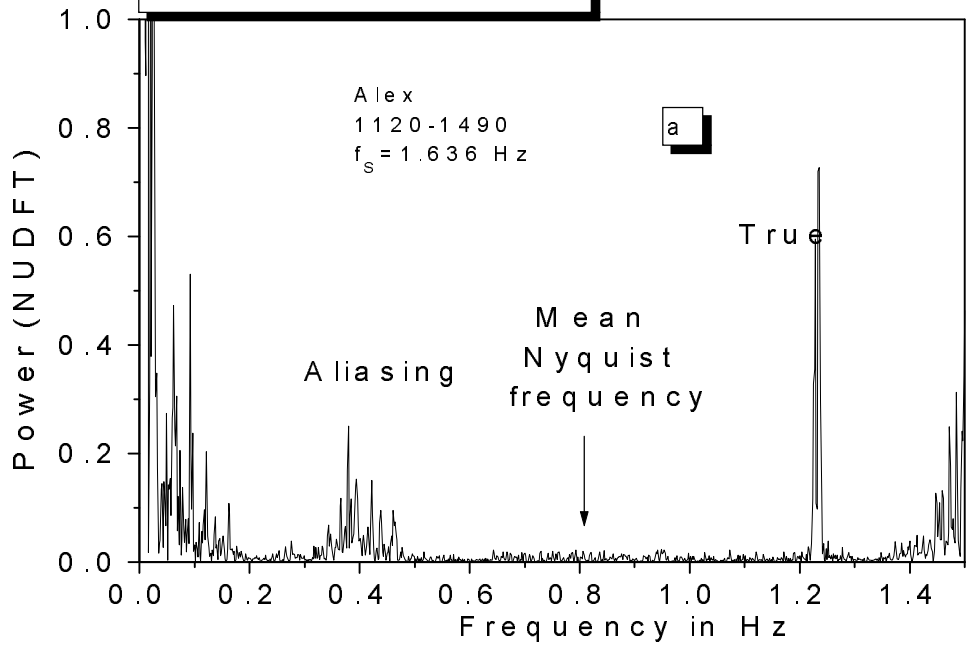
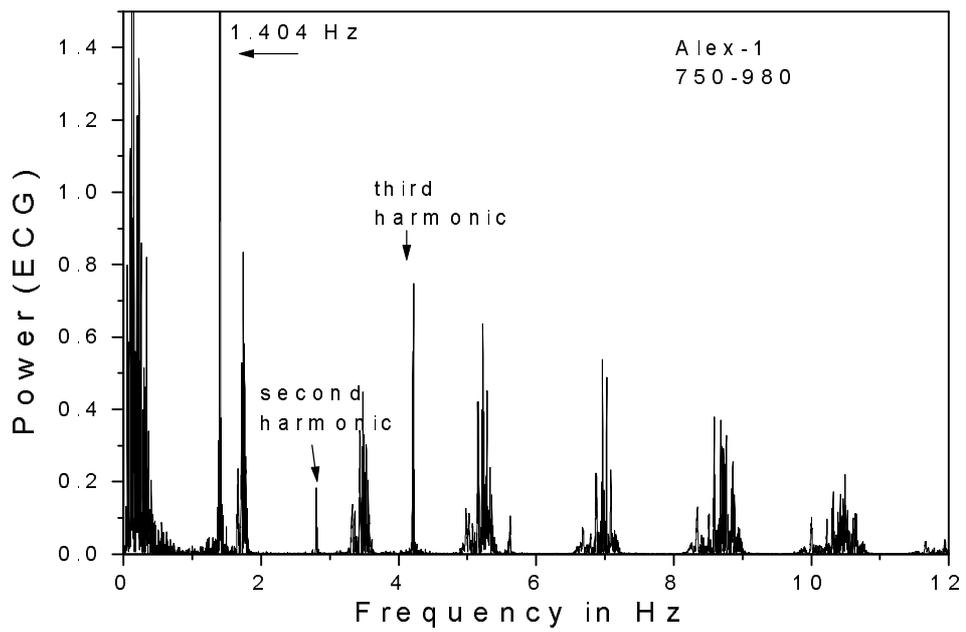
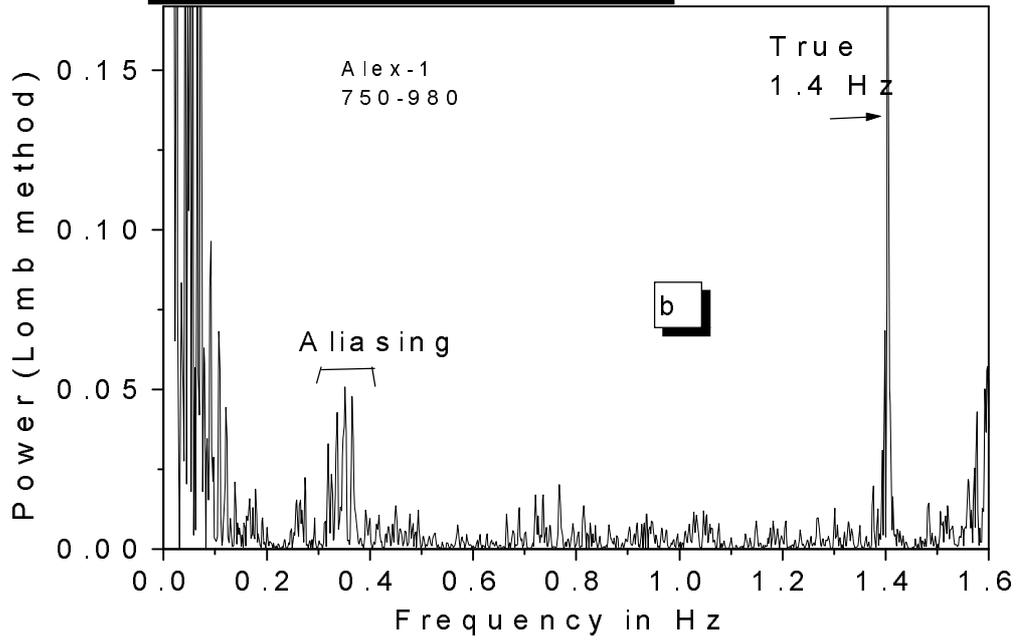
\end{document}